\documentclass[iop]{emulateapj}
 \slugcomment{{\sc Received 2013 January 20; Accepted 2013 February 11}}
\usepackage{natbib}
\usepackage{amsmath}

\newcommand{\bn}{\begin{enumerate}}
\newcommand{\en}{\end{enumerate}}
\newcommand{\bi}{\begin{itemize}}
\newcommand{\ei}{\end{itemize}}

\newcommand{\Msun}{M_\odot}

\newcommand{\himpc}{h^{-1} {\rm Mpc}}
\newcommand{\hikpc}{h^{-1} {\rm kpc}}

\newcommand{\Om}{\Omega_{\rm m}}

\newcommand{\Ob}{\Omega_{\rm b}}

\newcommand{\Muv}{M_{\rm uv}}

\newcommand{\Ebv}{E_{B-V}}

\newcommand{\Htwo}{\rm H_2}

\shorttitle{$\Htwo$-BASED STAR FORMATION MODEL, $z\ge 6$ LUMINOSITY FUNCTION, \& REIONIZTION}
\shortauthors{Jaacks, Thompson, \& Nagamine}

\begin{document}
\title{IMPACT OF $\Htwo$-BASED STAR FORMATION MODEL ON THE $z\geq6$ LUMINOSITY FUNCTION AND THE IONIZING PHOTON BUDGET FOR REIONIZATION}

\author{Jason Jaacks, Robert Thompson, Kentaro Nagamine\altaffilmark{\dag}}
\affil{Department of Physics \& Astronomy, University of Nevada, Las Vegas, 4505 S. Maryland Pkwy, Las Vegas, NV, 89154-4002, USA}
\email{jaacksj@physics.unlv.edu\vspace{0.2cm}}
\altaffiltext{\dag}{Visiting Scientist. Kavli Institute for the Physics and Mathematics for the Universe (IPMU), University of Tokyo, 5-1-5 Kashiwanoha Kashiwa, 277-8583, Japan}

%%%%%%%%%%%%%%%%%%%%%%%%%%%%%%%%%%%%%%%%%%%%%%%%%%%%%%%%%%%%%%%%%%%%%%

\begin{abstract}
We present the results of a numerical study examining the effect of $\Htwo$-based star formation (SF) model on the rest-frame UV luminosity function and star formation rate function (SFRF) of $z\geq6$ galaxies, and the implications for reionization.  Using cosmological hydrodynamical simulations outfitted with an $\Htwo$-SF model, we find good agreement with our previous results (non-$\Htwo$ SF model) and observations at $\Muv\leq-18$.  However at $\Muv>-18$, we find that the LF deviates from both our previous work and current observational extrapolations, producing significantly fewer low luminosity galaxies and exhibiting additional turnover at the faint end.  We constrain the redshift evolution of this turnover point using a modified Schechter function that includes additional terms to quantify the turnover magnitude ($\Muv^{t}$) and subsequent slope ($\beta$).  We find that $\Muv^t$ evolves from $\Muv^t=-17.33$ (at $z=8$) to $-15.38$ ($z=6$), while $\beta$ becomes shallower by $\Delta\beta=0.22$ during the same epoch.  This occurs in an $\Muv$ range which will be observable by \textit{James Webb Space Telescope}.  By integrating the SFRF, we determine that even though $\Htwo$-SF model significantly reduces the number density of low luminosity galaxies at $\Muv>-18$, it does not suppress the total SFR density enough to affect the capability of SF to maintain reionization. 
\end{abstract}


%%%%%%%%%%%%%%%%%%%%%%%%%%%%%%%%%%%%%%%%%%%%%%%%%%%%%%%%%%%%%%%%%%%%%%

\section{Introduction}
\label{sec:intro}

Characterizing the properties of galaxy population at high redshift is one of the most important goals of modern cosmological study, as star-forming galaxies are the primary candidate for the sources of ionizing photons that reionized the universe. 
In recent observational \citep[e.g.,][]{Bouwens.etal:12b, Finkelstein.etal:12,McLure.etal:12,Schenker.etal:12,Oesch.etal:12a} and theoretical \citep[e.g.,][]{Trenti.etal:10, Salvaterra.etal:11,Jaacks.etal:12a,Kuhlen.etal:12b} works, characteristics of the $z\geq6$ UV LF have been measured and predicted with increasing accuracy.  They both suggest that the faint-end slope of UV luminosity function (UV LF) is quite steep, and that the ionizing photon budget is dominated by the low-mass, star-forming galaxies below the current detection limit. 

In particular, some cosmological hydrodynamic simulations have found extremely steep faint-end slopes of $\alpha \lesssim -2.0$, which could be the result of too efficient star formation (SF) in their subgrid SF model at high-$z$. 
Previous SF models \citep[e.g.,][]{Springel:03, Schaye:08, Choi:10a} relied on the statistical representation of locally observed \citet{Kennicutt:98} relationship, which is dependent on the total gas column density. Recent observations \citep{Kennicutt.etal:07, Leroy.etal:08, Bigiel.etal:08} suggest however that SF correlates strongly with molecular hydrogen ($\Htwo$).  As the $\Htwo$ gas is less abundant at higher redshifts due to lower metallicities, the previous SF models that rely only on total gas density may be overproducing stars \citep{Krumholz.etal:09,Kuhlen.etal:12a}, potentially affecting the number density of low-mass galaxies at high-$z$. 

In this work, we quantify the effect of $\Htwo$-SF model on our simulated UV LF, star formation rate function (SFRF), and total SFR density (SFRD).  This paper is organized as follows: Section~\ref{sec:sim} briefly discusses the methods and parameters of our simulations;  Section~\ref{sec:results} presents our results, and we summarize in Section~\ref{sec:disc}. 

\section{Simulations}
\label{sec:sim}

\begin{deluxetable*}{ccccccc}
\tablecolumns{6}
\tablecaption{Simulation Parameters Used in this Paper. The parameter $N_p$ is the number of gas and dark matter particles; $m_{\rm DM}$ and $m_{\rm gas}$ are the particle masses of dark matter and gas; $\epsilon$ is the comoving gravitational softening length. }
\tablehead{
\colhead{Run}  & \colhead{Box Size} & \colhead{$N_{p}$} & \colhead{$m_{\rm{DM}}$} & \colhead{$m_{\rm gas}$} & \colhead{$\epsilon$} \\ 
\colhead{} & \colhead{($\himpc$)} & \colhead{(DM,Gas)}  & \colhead{($h^{-1} \Msun$)} & \colhead{($h^{-1} \Msun$)} & \colhead{($\hikpc$)}} \\
\startdata
N400L10 & $10.0$ & $400^{3}$ & $9.37{\times} 10^{5}$ & $1.91 {\times} 10^{5}$ & $1.0$  \\
N500L34 & $33.75$ & $500^{3}$ & $1.84 {\times} 10^{7}$ & $3.76 {\times} 10^{6}$ & $2.70$ \\
N600L100 & $100.0$ & $600^{3}$ & $2.78 {\times} 10^{8}$ & $5.65 {\times} 10^{7}$ & $4.30$  
\enddata
\label{tbl:Sim}
\end{deluxetable*}

We use a modified version of the smoothed particle hydrodynamics (SPH) code GADGET-3 \citep[originally described in][]{Springel:05}.  Our code includes radiative cooling by H, He, and metals \citep{Choi:09}, heating by a uniform UV background \citep[UVB;][]{Faucher.etal:09}, the UVB self-shielding effect \citep{Nagamine.etal:10}, the initial power spectrum of \citet{Eisenstein&Hu:99}, supernova feedback, sub-resolution multiphase interstellar medium model \citep{Springel:03}, and the Multicomponent Variable Velocity (MVV) wind model \citep{Choi:11a}.  Our current simulations do not include active galactic nucleus feedback.  

Simulations are setup with either $2\times 400^3$, $2\times 500^3$ or $2\times 600^3$ particles for both gas and dark matter. Multiple runs were made with comoving box sizes of $10h^{-1}$Mpc, $34h^{-1}$Mpc and $100h^{-1}$Mpc to cover a wide range of halo and galaxy masses.  Throughout this work they will be referred to as N400L10, N500L34 and N600L100 runs.  See Table~\ref{tbl:Sim} for simulation details.  The adopted cosmological parameters are consistent with the latest WMAP7 data: ${\Omega_{\rm m}}=0.26$, ${\Omega_{\Lambda}}=0.74$, ${\Omega_{\rm b}}=0.044$, $h=0.72$, $\sigma_{8}=0.80$, $n_{s} =0.96$  \citep{Komatsu:12}.  We note that the Fiducial runs used the \citet{Salpeter:55} stellar initial mass function (IMF), while the $\Htwo$ runs used the \citet{Chabrier:03} IMF for historical reasons of our work.   Galaxies are identified and grouped based on the baryonic density field \citep[see][for more details]{Nagamine:04}.

Since the estimation of $\Htwo$ mass fraction is dependent upon metallicity, the details regarding our feedback and enrichment models are relevant.  When SF takes place, metals are also produced with an instantaneous yield of $0.02$, and thereafter tracked by the code based on a closed box model for each gas particle (i.e., no diffusion). 
Our MVV wind model is designed to account for both energy-driven and momentum-driven winds \citep{Choi:11a}.  Wind velocity is determined by $v_{\rm wind}=\zeta v_{\rm esc}$, where $v_{\rm esc} = 130 (\rm{SFR})^{1/3} \left(\frac{1+z}{4}\right)^{1/2}$\,km\,s$^{-1}$.  We adopt the standard values of $\zeta=1.5$ for high-density regions (momentum driven) and $\zeta=1$ for low-density regions (energy driven), chosen by \citet{Choi:11a}.  The mass loading factor is $\eta = (\sigma_0 / \sigma_{\rm gal})^2$ for the energy-driven case, and $\eta = \sigma_0 / \sigma_{\rm gal}$ for the momentum-driven case, where $\sigma_0 = 300$\,km\,s$^{-1}$ and $\sigma_{\rm gal} = v_{\rm esc} / 2$ is the velocity dispersion of a galaxy.  For full detail and physical justifications for this model, see \citet{Choi:11a}.

Our "Fiducial" runs use the "Pressure-SF model" \citep{Schaye:08, Choi:10a}, while the present work uses the $\Htwo$-SF model of \citet{Krumholz.etal:09}, implemented by \citet[][hereafter $\Htwo$ runs]{Thompson.etal:13}.  This equilibrium analytic model calculates the SFR based on the $\Htwo$ mass density rather than the total cold gas density, and \citet{Krumholz&Gnedin:11} have shown that it is in good agreement with more computationally expensive, non-equilibrium calculations by \citet{Gnedin.etal:09}.  The details of the implementation and the basic results of this model have been presented by  \citet{Thompson.etal:13}.  

In principle our implementation of the $\Htwo$-SF model of \citet{Krumholz.etal:09} must be similar to the previous work by \cite{Kuhlen.etal:12a} on the most basic level.  The primary difference between the two work is in the class of code in which it was implemented, Enzo (AMR) versus GADGET (SPH).  We will further discuss the basic differences and potential effects in Sections~\ref{sec:lf} and \ref{sec:disc}.

%%%%%%%%%%%%%%%%%%%%%%%%%%%%%%%%%

\section{Results}
\label{sec:results}

\subsection{Modified Schechter Luminosity Function}
\label{sec:lf}

%%fig~1
\begin{figure*}
\begin{center}
\epsscale{1.2}
\plotone{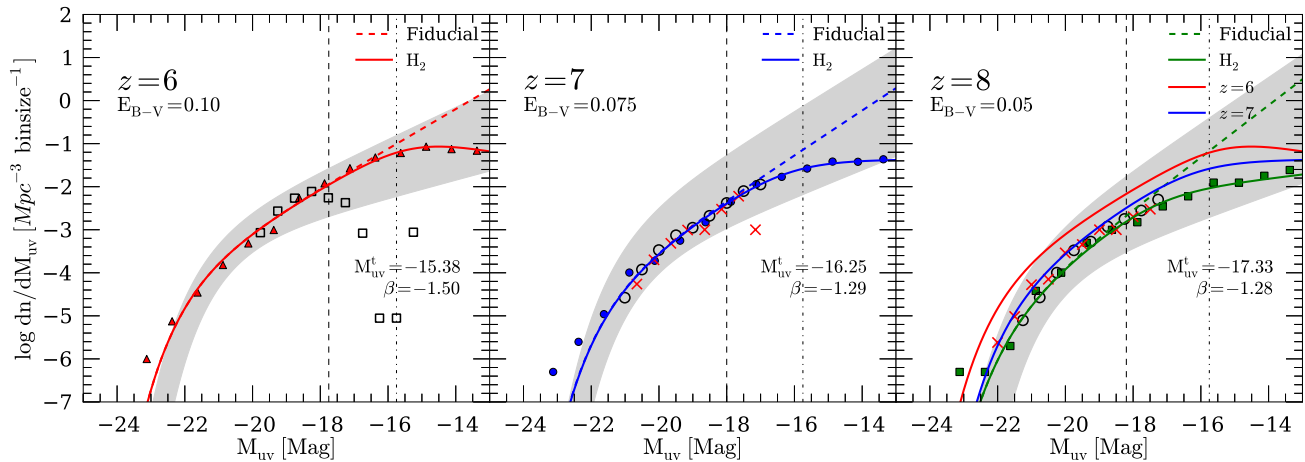}
\caption{Composite rest-frame UV LF at $z=6,7,8$ for the $\Htwo$ run, shown as red triangles, blue solid circles, and green triangles, respectively.  The solid lines represent the Schechter+ fit (Equation(\ref{schplus})) at each redshift to these simulation points. 
We also show the fits to the Fiducial run results (dashed lines) obtained by \citep{Jaacks.etal:12a}. 
The observed LF fit range \citep[gray shade;][]{Bouwens.etal:11a,Oesch.etal:12a} are also shown, which are extrapolated to $\Muv>-18$ based on their published fits.    Observed points from \citet{McLure.etal:12} and \citet{Schenker.etal:12} are shown by the open black circles and red crosses at $z=7$ \& 8, respectively.  The vertical dashed line indicates the approximate \textit{HST} observational limit ($\Muv\sim-18$) at these redshifts and the vertical dash-dotted line represents the approximate \textit{JWST} limit ($\Muv\sim-16$).  The $z=8$ panel also includes $z=6\ \&\ 7$ line (red \& blue) to demonstrate evolution.
We show the results of Enzo AMR simulation from \citet[][black open squares]{Kuhlen.etal:12a} in the $z=6$ panel for direct comparison. 
}
\label{fig:LFsch}
\end{center}
\end{figure*}

We combine the results of our three runs to create a composite LF, which covers a much wider dynamic range than is possible with a single cosmological run.  In Figure~\ref{fig:LFsch}, we present our composite LF for $z=6, 7, 8$ (red triangles, blue circles, green squares) for the $\Htwo$ run,  in comparison to the \citet{Schechter:76} fits for our Fiducial runs \citep[dashed red, blue, green lines;][]{Jaacks.etal:12a} with the Pressure-SF model.  We also show the observed LF fit range \citep[gray shade;][]{Bouwens.etal:11a}.  A small, constant extinction $\Ebv$ is required to fall within the observational range for both runs \citep{Jaacks.etal:12a}, although the $\Htwo$ runs at $z=7$ \& 8 require less extinction by $\Delta \Ebv=0.025$ than the Fiducial runs,  suggesting a trend of decreasing $\Ebv$ with increasing redshift. 

The value of  $E_{B-V}$ is chosen to be consistent with the value used to match the observed rest-frame UV LF in our previous work \citep{Jaacks.etal:12a}, and it is centered between the following two recent observations: 
\citet{Bouwens.etal:12a} argued for little to no extinction at the faint end of the LF at $z=6$, whereas  
\citet{Willott.etal:12} found a best-fit value of $A_V=0.75$, which corresponds to $E_{B-V}\sim 0.19$ assuming $R_V=4.05$ \citep{Calzetti.etal:00} at the bright end of the UV LF at $z=6$.  This moderate amount of extinction is also consistent with the estimates by \citet{Schaerer.deBarros:10} and \citet{deBarros.etal:12} who included nebular emission lines in their spectral energy distribution fits. 
Therefore the values of $E_{B-V}$ chosen for this work are reasonably consistent with current observations. 

At $\Muv \lesssim -18$, both Fiducial and $\Htwo$ runs show excellent agreement with each other and observations.  However at $\Muv > -18$, the $\Htwo$ run start to show a turn-over of the LF, which is not present in the Fiducial LFs.  This flattening significantly reduces the number density of low luminosity objects in the $\Htwo$ run, and it occurs beyond current observational limit of the \textit{Hubble Space Telescope} (\textit{HST}).  As this population of low-luminosity galaxies is thought to be the critical contributor to the total ionizing flux at these redshifts \citep{Trenti.etal:10, Salvaterra.etal:11, Bouwens.etal:12b, Finkelstein.etal:12,Jaacks.etal:12a}, it is important to quantify this reduction and its implications.

To quantify the turnover point and flattening, we adopt a modified Schechter function (hereafter Schechter+): 
\begin{equation}
\Phi(L)=\phi^*\left(\frac{L}{L^*}\right)^\alpha \exp \left(-\frac{L}{L^*}\right)\left[1+\left(\frac{L}{L^t}\right)^\beta\right]^{-1},
\label{schplus}
\end{equation}
where $\phi^{*}$, $L^{*}$ and $\alpha$ are the normalization, characteristic luminosity, and faint-end slope of the standard Schechter function.  
The additional parameter $L^t$ indicates the point at which the LF undergoes its second turn, and $\beta$ is related to the power-law slope at the lowest luminosities. 
Note that $\Phi(L) \propto L^{\alpha-\beta}$ when $L \ll L^t$, 
and that both $\alpha$ and $\beta$ take negative values. 
A similar functional form to Equation~(\ref{schplus}) was used by \citet{Loveday:97} to characterize the upturn in dwarf galaxy population at low redshift. 

Using the magnitude version of Equation~(\ref{schplus}), we perform a least-square fit to the $\Htwo$ runs by fixing the three standard Schechter parameters ($\phi^*,\Muv^*,\alpha$) to those obtained for the Fiducial runs by \citet{Jaacks.etal:12a}, and vary the remaining two parameters ($\Muv^t,\beta$).  
The best-fit results can be seen as solid red, blue and green lines in Figure~\ref{fig:LFsch} and are quantified in Table~\ref{tbl:sch}.

In the rightmost $z=8$ panel, we have overplotted the fits to $z=6$ \& 7 results to demonstrate the evolution.  
One can clearly see that the number of galaxies increases from $z=8$ to $z=6$ as the dark matter structure develops and more SF takes place. 

One might note the somewhat large uncertainty in the estimation of $\beta$.  This is due to the fact that we are fixing the standard Schechter parameters ($\phi^*, \Muv^*, \alpha$) to our Fiducial results and only fitting the new Schechter+ parameters ($\Muv^t, \beta$).  This was done in order to make a valid comparison between the current results and our previous results \citep{Jaacks.etal:12a}.  If we make a full five parameter fit, we noticed that the value of $\Muv^*$ becomes unphysically brighter and $\alpha$ becomes quite steeper, due to the chosen Schechter+ functional form.  As a result, the size of the error bar of $\beta$ becomes smaller, however there would be no meaningful physical comparison to be made between the three parameter Schechter values which was obtained in our previous work. 
Therefore we chose to fix the standard Schechter parameters as in our previous work, despite of large error bar on $\beta$.  

\begin{deluxetable}{ccccc}
\tablecolumns{5}
\tablecaption{Best-fit Schechter+ LF Parameters at $z=6-8$ (Top to Bottom) for the $\Htwo$ Runs. 
The errors are 1$\sigma$.  
 }
\tablehead{
\colhead{$\alpha$} & \colhead{$\log( \phi^*)$} & \colhead{$\Muv^*$} & \colhead{$\Muv^t$} & \colhead{$\beta$}  \\ 
 \colhead{(Fixed)}	     & \colhead{(Fixed)}	& \colhead{(Fixed)} 	 &	\colhead{}	    &	\colhead{}	\vspace{0.1cm}}\\  

\startdata
$-2.15^{+.24}_{-.15}$ & $-3.46^{+.29}_{-.37}$ & $-21.15^{+.53}_{-.53}$ & $-15.38^{+.60}_{-.60}$&$-1.50^{+.47}_{-.47}$	\vspace{0.2cm} \\
$-2.30^{+.28}_{-.18}$ & $-3.74^{+.32}_{-.41}$ & $-20.82^{+.61}_{-.56}$ & $-16.25^{+1.33}_{-1.33}$&$-1.29^{+.70}_{-.70}$	\vspace{0.2cm} \\
$-2.51^{+.27}_{-.17}$ & $-4.30^{+.32}_{-.42}$ & $-21.00^{+.59}_{-.50}$ & $-17.33^{+1.21}_{-1.21}$&$-1.28^{+.49}_{-.49}$	
%\vspace{0.2cm} \\
\enddata
\label{tbl:sch}
\end{deluxetable}

In the leftmost $z=6$ panel, we have over plotted the Enzo AMR simulation from \citet[][black open squares, corrected for dust extinction by the same amount as our simulation]{Kuhlen.etal:12a}, which also used the same \citet{Krumholz.etal:09} $\Htwo$-SF model. At the brighter side of $\Muv<-18$, we see good agreement between the two simulations.  However, on the fainter side of $\Muv > -18$, the Enzo result drops sharply relative to our SPH results.  While the origin of this difference is not entirely clear, one possible explanation is that the AMR simulation is missing the early structure formation of low-mass dark matter halos relative to our SPH simulations \citep{Oshea.etal:05}. The simulation of \citet{Kuhlen.etal:12a} used a comoving box size of 12.5 Mpc, which is not very different from our N400L10 run.  However, their root grid is only 256$^3$, which effectively corresponds to 128$^3$ particles of SPH/$N$-body simulation \citep{Oshea.etal:05}.  Although their highest spatial resolution after recursive refinement might be higher than our simulation, their AMR simulation would miss a large number of low-mass halos at early times due to lack of force resolution at early times, hence significantly lower number of dwarf galaxies compared to our SPH simulations which used $400^3$ dark matter particles.  This is a well-known issue of AMR cosmological simulation, and has been extensively discussed by \citet{Oshea.etal:05}.  If \citet{Kuhlen.etal:12a} wanted to resolve the low-mass halos down to the same halo mass as our N400L10 run, they had to start their simulation with $\sim 800^3$ root grid and correspondingly low dark matter particle mass.  In their Figure~1, \citet{Kuhlen.etal:12a} show that their halo mass function starts to deviate from \citet{Sheth.Torman:99} at $M_{\rm halo} \lesssim 10^9 \Msun$, whereas our halo mass function in N400L10 run agrees very well with \citet{Sheth.Torman:99} down to $10^8 \Msun$, as we showed in Figure~17 of \citet{Jaacks.etal:12a}.  This corroborates our suspicion that the main difference in the two results is coming from the difference in the number of low-mass halos in the two simulations.  We also discuss the other possibility of difference in metal diffusion in Section~\ref{sec:disc}. 

\begin{figure}
\begin{center}
\includegraphics[scale=0.44] {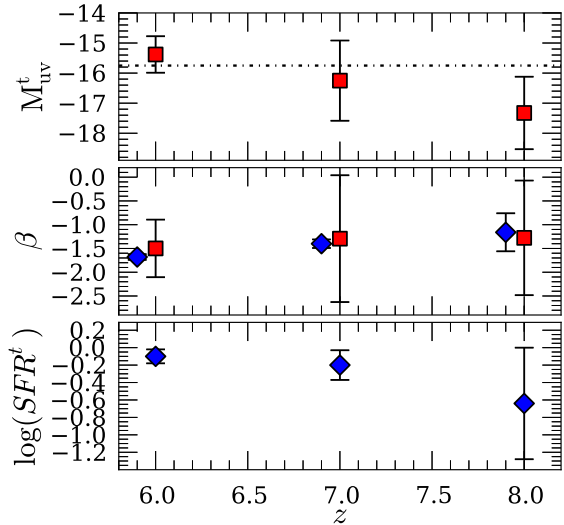}
\caption{Redshift evolution of Schechter+ parameters for both our simulated LF (red squares) and simulated SFRF (blue diamonds) at $z=6, 7,$ \& 8, shown with 1$\sigma$ error bars.  The SFRF $\beta$ value is offset by $\Delta z=0.1$ for readability.  The horizontal dash-dotted line in the top panel represents the approximate \textit{JWST} observable limit ($\Muv \approx -16$).
}
\label{fig:params}
\end{center}
\end{figure}

In Figure~\ref{fig:params}, we summarize the redshift evolution of best-fit Schechter+ parameters. 
We see that $\Muv^t$ becomes dimmer and the power-law slope $\alpha - \beta$ of $\Phi(L)$ becomes shallower with decreasing redshift.  These results support the current paradigm of hierarchical structure formation: as the smaller objects form first and later merge to form larger systems, the number of dim object decreases, and the LF becomes flatter at the faintest end.  Since the predicted $\Muv^t$ occurs at $\Muv<-16$ at $z=7$ \& 8, this feature should be observable by future missions such as \textit{James Webb Space Telescope} (\textit{JWST}).
 
%%fig~2

%%%%%%%%%%%%%%%%%%%%%

\subsection{Star Formation Rate Function (SFRF)}
\label{sec:sfrf}
A recent work by \citet{Smit.etal:12} presented the observed SFRF from dust-corrected UV LF by 
converting $L_{\rm{UV}}$ into SFR using the \citet{Kennicutt:98} conversion.  The SFRD is then easily obtained by integrating the Schechter-like fit to the SFRF down to a certain lower limit, culminating in the commonly used Lilly-Madau diagram \citep{Lilly.etal:96, Madau.etal:96}.  Comparing the SFRF of our simulated galaxies to the observed results is an excellent test of our SF model, since SFR is one of the most basic outputs of our simulations.  It is also a more intrinsic comparison for us as it does not require any assumptions on our part regarding the amount of extinction in our simulated galaxies as this correction is included in the observational estimates. 

\begin{figure*}
\begin{center}
\epsscale{1.2}
\plotone{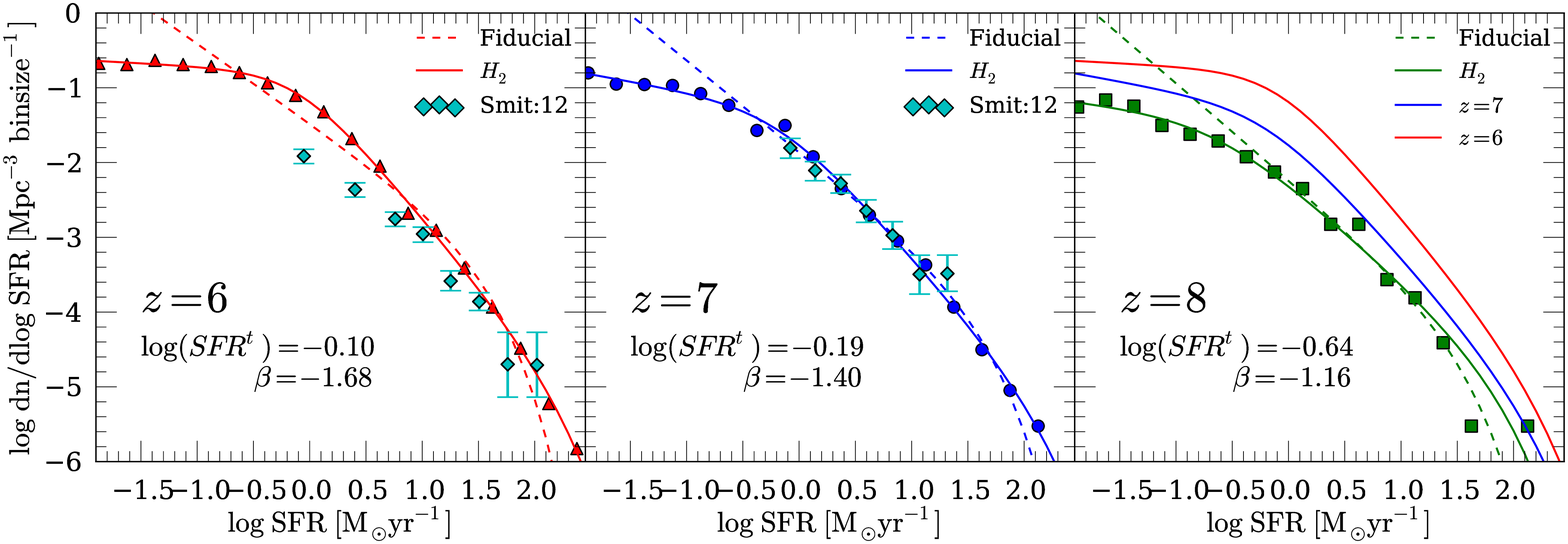}
\caption{SFRF of simulated galaxies at $z=6, 7,$\,\&\,8, shown as red triangles, blue circles and green squares, respectively.  The observational estimates \citep{Smit.etal:12} are shown by filled cyan diamonds. 
Solid red, blue and green lines represent the best-fit Schechter+ functions (Equation~\ref{eq:sch_sfrf}) to the simulation data. 
}
\label{fig:sfrf}
\end{center}
\end{figure*}

Figure~\ref{fig:sfrf} shows the SFRF of simulated galaxies at $z=6, 7, 8$ (red triangles, blue circles and green squares, respectively).  We find good agreement with the observed results \citep[][cyan diamonds]{Smit.etal:12}, especially at $z=7$.  Deviation at low-SFR end of $z=6$ is expected, given a similar deviation in the LF at the low-end shown in \citet{Jaacks.etal:12a,Jaacks.etal:12b}.  
This deviation could either indicate that our simulation is still overproducing stars at $z=6$, or it could also be due to uncertainties in the faint-end observations (i.e., missed faint galaxies) and/or assumptions regarding extinction.  

The SFRF has a similar functional form as the LF, therefore we utilize the same Schechter+ function to fit it: $\phi(\rm{SFR})\equiv dn/d \log({\rm SFR}) = \ln(10) {\rm SFR}\ \Phi({\rm SFR})$, and  
\begin{align}
\nonumber \phi({\rm SFR})=\ln(10)\phi^*\left(\frac{\rm SFR}{\rm SFR^*}\right)^{(1+\alpha)}\exp\left(-\frac{\rm SFR}{\rm SFR^*}\right)\\
\times \left[1+\left(\frac{\rm SFR}{\rm SFR^t}\right)^\beta\right]^{-1},
\label{eq:sch_sfrf}
\end{align}
where $\phi^{*}$, SFR$^{*}$ and $\alpha$ are the usual Schechter parameters as in Equation~(\ref{schplus}); SFR$^t$ is the location of the second turnover, and $\beta$ is related to the low-SFR end power-law slope which is proportional to ($1+\alpha-\beta$) at ${\rm SFR} \ll {\rm SFR}^t$.  
The best-fit results to all five parameters are shown in Figure~\ref{fig:sfrf} with solid lines. 
Similarly to the rest-frame UV LF, the right-most panel shows the increase in normalization and brightening of SFR$^t$ 
clearly from $z=8$ to $z=6$. 

%%fig~3

%%%%%%%%%%%%%%%%%%%%%%
%%fig~4
\begin{figure}
\begin{center}
\includegraphics[scale=0.35] {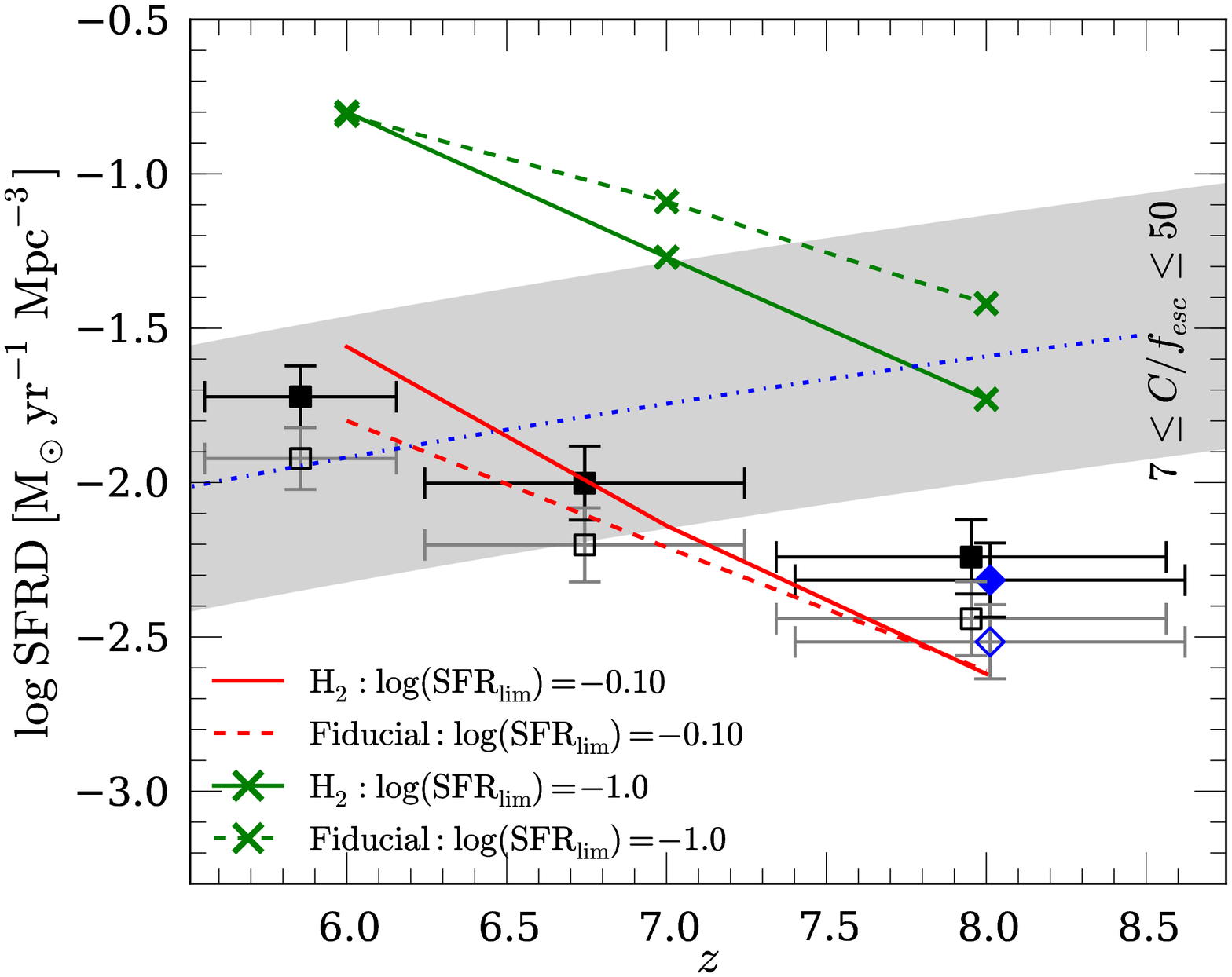}
\caption{SFRD obtained from the integration of SFRF for both $\Htwo$ and Fiducial runs with SFR limits of $\log({\rm SFR})=-0.10$ (solid/dashed red line) and $\log({\rm SFR})=-1.0$ (solid/dashed green crossed line).  Observations from \citet{Bouwens.etal:12a} are represented by the black squares and recent $z=8$ observations from \citet{Oesch.etal:12a} by a blue diamond.  The open black squares and open blue diamond show the same observed points adjusted for a Chabrier IMF.  The SFRD required to maintain IGM ionization (Equation~\ref{madau}) is shown by the shaded gray area for $7\leq C/f_{\rm esc} \leq 50$.  The blue dot-dashed line represents an updated estimation of the critical SFRD by \citet{Shull.etal:12}.
}
\label{fig:sfrd}
\end{center}
\end{figure}

\subsection{SFRD and Reionization}
\label{sec:reion}

Having a continuous representation of the SFRF allows for easy integration and acquisition of the SFRD at each redshift.  In Figure~\ref{fig:sfrd} we show the results of this integration (red solid and green lines) plotted against observational results by \citet[][black squares]{Bouwens.etal:11a} and \citet[][blue diamond]{Oesch.etal:12a}.  
The red solid line represents an integration down to $\log({\rm SFR}_{\rm lim}[\Msun {\rm yr}^{-1}])=0.0$, which is approximately consistent with the lowest SFR value of the observational data.  The green solid crossed line is the SFRD value when integrated down to $\log({\rm SFR}_{\rm lim})\approx-1.0$.  This value is chosen to be consistent with a galaxy stellar mass $M_s=10^7 \Msun$ via the $M_s$-SFR relationship found in our simulations (SFR$\propto M_s^{1.0}$).  The mass of  $M_s=10^7 \Msun$ also represents the minimum galaxy resolution of the N400L10 run ($\sim$100 star particles).  The solid observational data points were obtained assuming the Salpeter IMF, therefore we also show the same data points corrected for the Chabrier IMF with open symbols by applying a simple factor \citep[see Section 3.2.1 of][]{Thompson.etal:13}.  Our conclusions are not affected by the change in the IMF. 

Figure~\ref{fig:sfrd} demonstrates that, when integrated down to a minimum SFR ($\log(SFR_{\rm lim})= -0.10$), the $\Htwo$ run agrees very well with the Fiducial run and observational estimates.  This is expected, since the $\Htwo$-SF model did not affect the number density of objects within the observed range of $\Muv\lesssim -18$ (Figure~\ref{fig:LFsch}).  We note that the observed points were calculated by integrating the UV LF  to a limiting $\Muv\sim-17.74$ \citet{Bouwens.etal:12a}, which corresponds to $\log({\rm SFR})\approx -0.10$ when converted via \citet{Kennicutt:98} relation. 

The green crossed line, representing the integration limit of $\log({\rm SFR}_{\rm lim})= -1.0$, shows that the total SFRD in our simulations is significantly higher than what is currently observed.  This is consistent with our previous findings \citep{Choi.etal:12, Jaacks.etal:12a}.  The reduction of SFRD at $z\geq6$ is also consistent with findings by \citet{Krumholz.etal:09, Gnedin.etal:10}, and \citet{Kuhlen.etal:12a}, who show that $\Htwo$-SF model reduces high-$z$ SFRD due to metallicity effect \citep{Thompson.etal:13}.   
However the degree of reduction may still be different among different simulations and models. 

To determine whether or not our simulated galaxy population is sufficient to maintain reionization, we utilize the theoretical prescription presented in \citet{Madau.etal:99}, which quantifies the minimum SFRD required to keep the intergalactic medium (IGM) ionized (shaded gray contour):
\begin{equation} 
\label{madau}
\dot\rho_\star\approx2\times10^{-3}\left( \frac{C}{f_{\rm esc}}\right ) \left(\frac{1+z}{10}\right)^{3}\ [M_{\odot}\,{\rm yr}^{-1}{\rm Mpc}^{-3}]. 
\end{equation}
This depends on redshift and the ratio of IGM clumping factor ($C$) and escape fraction ($f_{\rm esc}$) of ionizing photons from galaxies.  Given that the exact values of both $C$ and $f_{\rm esc}$ are still debated and uncertain, we show a wide range of $7\leq C/f_{\rm esc} \leq 50$ (grey shade in Figure~\ref{fig:sfrd}) which are consistent with works by \citet{Iliev.etal:06,Pawlik.etal:09,Finlator.etal:12,Kuhlen.etal:12b}.

We also include an updated estimation of the critical SFRD by \citet[][blue dot-dashed line]{Shull.etal:12} which includes considerations for the Ly-continuum production rates and the temperature scaling of the recombination rate coefficient. Their calculation uses a fiducial value of $C/f_{\rm esc}=15$.

Figure~\ref{fig:sfrd} reinforces our previous arguments \citep{Jaacks.etal:12a} that the abundant, low-luminosity galaxies, which are currently beyond the detection threshold of \textit{HST}, dominate the total SFRD at $z\geq6$.  When this population is considered, there is sufficient amount of ionizing photons available to maintain ionization by $z=6-7$ for a reasonable value of $C/f_{\rm esc}$.

%%%%%%%%%%%%%%%%%%%%%%%%%%%

\section{Conclusions and Discussions}
\label{sec:disc}
Using {\small GADGET-3} cosmological SPH simulations equipped with a $\Htwo$-SF model, we examined UV LF, SFRF and the contribution of low-luminosity galaxies to the total SFRD at $z\geq6$ .  Our major conclusions are as follows. 
\begin{itemize}
\item We find that, at $\Muv \lesssim-18$, the $\Htwo$-based SF model does not change the faint-end slope from our Fiducial runs with $\alpha<-2.00$ at $z\geq6$  \citep{Jaacks.etal:12a}.  However at the fainter end of $\Muv > -18$, we find a significant reduction in number density of faint galaxies due to the metallicity effect of $\Htwo$-SF model \citep{Thompson.etal:13}. 

\vspace{0.2cm}
\item  To characterize the new LF's turnover magnitude ($\Muv^t$) and subsequent slope ($\beta$), we utilize the Schechter+ function (Equation~\ref{schplus}), and find that $\Muv^t$ evolves with redshift from $\Muv^t=-15.33$ ($z=6$) to $\Muv^t=-17.69$ ($z=8$), 
while the second power-law slope ($1+\alpha - \beta$) becomes steeper over the same redshift range with $\Delta\beta=0.22$ (Figure~\ref{fig:params}). 

\vspace{0.2cm}
\item  We present the SFRF of simulated galaxies (Figure~\ref{fig:sfrf}), and find good agreement with observations.  We also find similar evolution in the Schechter+ parameters of SFRF: as $\log({\rm SFR}^t)$ becomes slightly lower with increasing redshift, and the second power-law ($1+\alpha - \beta$) becomes steeper over the same redshift with $\Delta\beta=0.52$ (Figure~\ref{fig:params}).  

\vspace{0.2cm}
\item Although the $\Htwo$-SF model reduces the number density of low-luminosity galaxies, there is still sufficient SFRD (i.e., ionizing photons) to maintain reionization at $6\leq z \leq8$ (Figure~\ref{fig:sfrd}) for a wide parameter range of clumping factor ($C$) and escape fraction ($f_{\rm esc}$).

\end{itemize}

When examining the results of numerical simulations, the resolution effect is always a concern.  The N400L10 run used in the present work is reaching the densities similar to that of giant molecular clouds (GMCs), but it is not resolving the formation and internal structure of GMCs adequately.  The model of \citet{Krumholz.etal:09} is meant to capture the general behavior of GMCs on average, and \citet{Thompson.etal:13} have shown that the $\Htwo$-SF model has a number of advantages over the previous SF models.  Nevertheless, as discussed by \citet{Thompson.etal:13}, the $\Htwo$-SF model has some inherent resolution dependency.  Although we believe that the observed turnover in the UV LF and SFRF are physical in nature, the exact values of $\Muv^{t}$ and $\log({\rm SFR}^t)$ may vary with resolution. 

Another concern is the impact of UVB on the faint-end slope of LF. 
Earlier work using semi-analytic models \citep[e.g.,][]{Babul.etal:92,Bullock.etal:00,Benson.etal:03} and simulations \citep[e.g.,][]{Thoul.etal:96,Gnedin:00, Hoeft.etal:06, Okamoto.etal:08,Finlator.etal:11b} have shown that the UVB suppresses the formation of low-mass galaxies via photoionization of gas in shallow potential wells.  
In particular, \citet{Okamoto.etal:08} have shown that the gas fraction drops rapidly from cosmic fraction ($\Ob/\Om$) at $M_{\rm tot}$$\sim$$10^9 \Msun$ to almost zero at $M_{\rm tot}$$\sim$$10^8 \Msun$. 
It is therefore plausible that the turnover seen in our LF and SFRF could be the result of photoevaporation of gas from low-mass halos and not a feature of the implemented $\Htwo$-SF model.  In order to verify that the turnover at $\Muv^t$ and ${\rm SFR}^t$ in the $\Htwo$ runs are not affected by the UVB, we ran two test simulations with and without UVB. As a result, we found that the turn-over still exists even in the no-UVB run, and only the  normalization of UV LF decreased slightly due to photoionization.  The overall shape of the LF and SFRF remained unchanged, i.e., the values of $\Muv^t$ and ${\rm SFR}^t$ did not change significantly.  Therefore, while the UVB reduces the number of low-mass galaxies in our simulations, we confirmed that the turn-over at $\Muv^t$ and SFR$^t$ is due to the $\Htwo$-SF model. 

In Section~\ref{sec:lf} we discussed the differences between the results of Enzo AMR simulation and our GADGET SPH simulations. 
Another fundamental difference between the two codes is in metal mixing, which could have an effect on these results.  In our SPH code metals are locked to the individual SPH particles, thus limiting mixing and diffusion of metals.  In contrast grid-based codes follow the fluid element in cells, which may allow for more natural mixing and diffusion within and between the cells.  Conversely in low resolution regions, this could lead to over-smoothing of metals.  At the moment, it is unclear how much impact the metal mixing and diffusion would have on the results presented in this paper, unless we perform more rigorous comparison tests between the two codes.  We intend to investigate  the issue of metal mixing and diffusion in the future using both SPH and AMR codes.

%%%%%%%%%%%%%%%%%%%%%%%%%%%%%%%%%%%%%%%%%%

\section*{Acknowledgments}

We are grateful to V. Springel for allowing us to use the original version of {\small GADGET-3} code.  We also thank Kristian Finlator for fruitful discussions, and Jun-Hwan Choi for some of our Fiducial simulations.  This work is supported in part by the NSF grant AST-0807491, and National Aeronautics and Space Administration under Grant/Cooperative Agreement No. NNX08AE57A issued by the Nevada NASA EPSCoR program.  Support for Program number HST-AR-12143-01-A was provided by NASA through a grant from the Space Telescope Science Institute, which is operated by the Association of Universities for Research in Astronomy, Incorporated, under NASA contract NAS5-26555.  This research is also supported by the NSF through the TeraGrid/XSEDE resources provided by the Texas Advanced Computing Center and the National Institute for Computational Sciences. 
KN acknowledges the hospitality of the Kavli Institute for Physics and Mathematics of the Universe (IPMU), University of Tokyo, the Aspen Center for Physics, and the National Science Foundation Grant No. 1066293.

%%%%%%%%%%%%%%%%%%%%%%%%%%%%%%%%%%%%%%%%%%%

%\bibliographystyle{mn2e}
\bibliographystyle{apj}

%\bibliography{jjref}

\end{document}